\documentclass[sigconf]{acmart}

\AtBeginDocument{%
  \providecommand\BibTeX{{%
    \normalfont B\kern-0.5em{\scshape i\kern-0.25em b}\kern-0.8em\TeX}}}

\setcopyright{rightsretained}
\copyrightyear{2021} 
\acmYear{2021} 
\setcopyright{rightsretained} 
\acmConference[CHI '21]{CHI Conference on Human Factors in Computing Systems}{May 8--13, 2021}{Yokohama, Japan}
\acmBooktitle{CHI Conference on Human Factors in Computing Systems (CHI '21), May 8--13, 2021, Yokohama, Japan}\acmDOI{10.1145/3411764.3445326}
\acmISBN{978-1-4503-8096-6/21/05}

\usepackage{alltt}
\usepackage{fixfoot}

\DeclareFixedFootnote\Bootstrap{Bootstrap examples: \url{https://getbootstrap.com/docs/4.0/examples/}}

\newif\ifsubmit
\submitfalse
\ifsubmit
    \newcommand{\mingyuanz}[1]{}
    \newcommand{\leebird}[1]{}
    \newcommand{\liyang}[1]{}
    \newcommand{\peggy}[1]{}
\else
    \newcommand{\mingyuanz}[1]{{\textbf{\color{blue}{Mingyuan: {#1}}}}}
    \newcommand{\leebird}[1]{{\textbf{\color{green}Gang: {#1}}}}
    \newcommand{\liyang}[1]{{\textbf{\color{red}{Yang: {#1}}}}}
    \newcommand{\peggy}[1]{{\textbf{\color{orange}Peggy: {#1}}}}
\fi

\begin{document}

\title{Spacewalker: Rapid UI Design Exploration Using Lightweight Markup Enhancement and Crowd Genetic Programming}

\author{Mingyuan Zhong}
\authornote{This work was completed while the author was an intern at Google Research.}
\affiliation{%
  \institution{University of Washington}
  \city{Seattle}
  \state{WA}
}
\email{myzhong@cs.washington.edu}

\author{Gang Li}
\affiliation{%
  \institution{Google Research}
  \city{Mountain View}
  \state{CA}
}
\email{leebird@google.com}

\author{Yang Li}
\affiliation{%
  \institution{Google Research}
  \city{Mountain View}
  \state{CA}
}
\email{liyang@google.com}









\begin{abstract}
  User interface design is a complex task that involves designers examining a wide range of options. We present Spacewalker, a tool that allows designers to rapidly search a large design space for an optimal web UI with integrated support. Designers first annotate each attribute they want to explore in a typical HTML page, using a simple markup extension we designed. Spacewalker then parses the annotated HTML specification, and intelligently generates and distributes various configurations of the web UI to crowd workers for evaluation. We enhanced a genetic algorithm to accommodate crowd worker responses from pairwise comparison of UI designs, which is crucial for obtaining reliable feedback. Based on our experiments, Spacewalker allows designers to effectively search a large design space of a UI, using the language they are familiar with, and improve their design rapidly at a minimal cost. 
\end{abstract}



\maketitle

\section{Introduction}
User interface design is a complex task that often requires designers to explore a wide range of options, which is expensive and time consuming. For example, a designer may consider multiple color schemes or layout choices for a UI. To evaluate these options, it is often necessary to test them with users, via either usability testing \cite{dumas1999practical, corry1997user} or A/B testing at scale \cite{Kohavi2017, ab_oreilly}. Although these classical approaches are widely used, they require substantial engineering investment to build and instrument each design alternative for testing, and extensive analytical effort to process collected user data and distill findings.

To ease the effort for exploring a design space, previous work has extensively investigated using crowdsourcing as an essential component in UI design and evaluation \cite{lasecki2015apparition, lee2018exploring, park2013crowd, lee2017sketchexpress, reinecke2014quantifying, deka2017zipt, xu2014voyant, oppenlaender2020crowdui, wang2019kaleidoscope}, which lowers the threshold for acquiring user feedback at scale. Various commercial tools also exist to support A/B testing of UI designs. Nevertheless, it remains challenging for existing tools to examine a large design space where it is a commonplace to have hundreds or even thousands of design alternatives.

To battle the issue, previous work has attempted to apply Artificial Intelligence algorithms to enable efficient search of a large design space \cite{monmarche1999imagine, quiroz2007interactive, salem2017user, tamburrelli2014towards, iitsuka2015website}. Particularly, Salem \cite{salem2017user} combined crowdsourcing and genetic programming \cite{10.5555/522098} for the design of landing pages. However, these existing tools often require a designer to learn a new language that is tailored for working with the underlying algorithm to define a search space. Their optimization objectives (or fitness functions \cite{10.5555/522098}) are designed based on user click behaviors of specific interaction tasks. There lacks a tool for general-purpose design space exploration that is seamlessly integrated into current design practice.

\begin{figure*}
  \centering
  \includegraphics[width=0.95\linewidth]{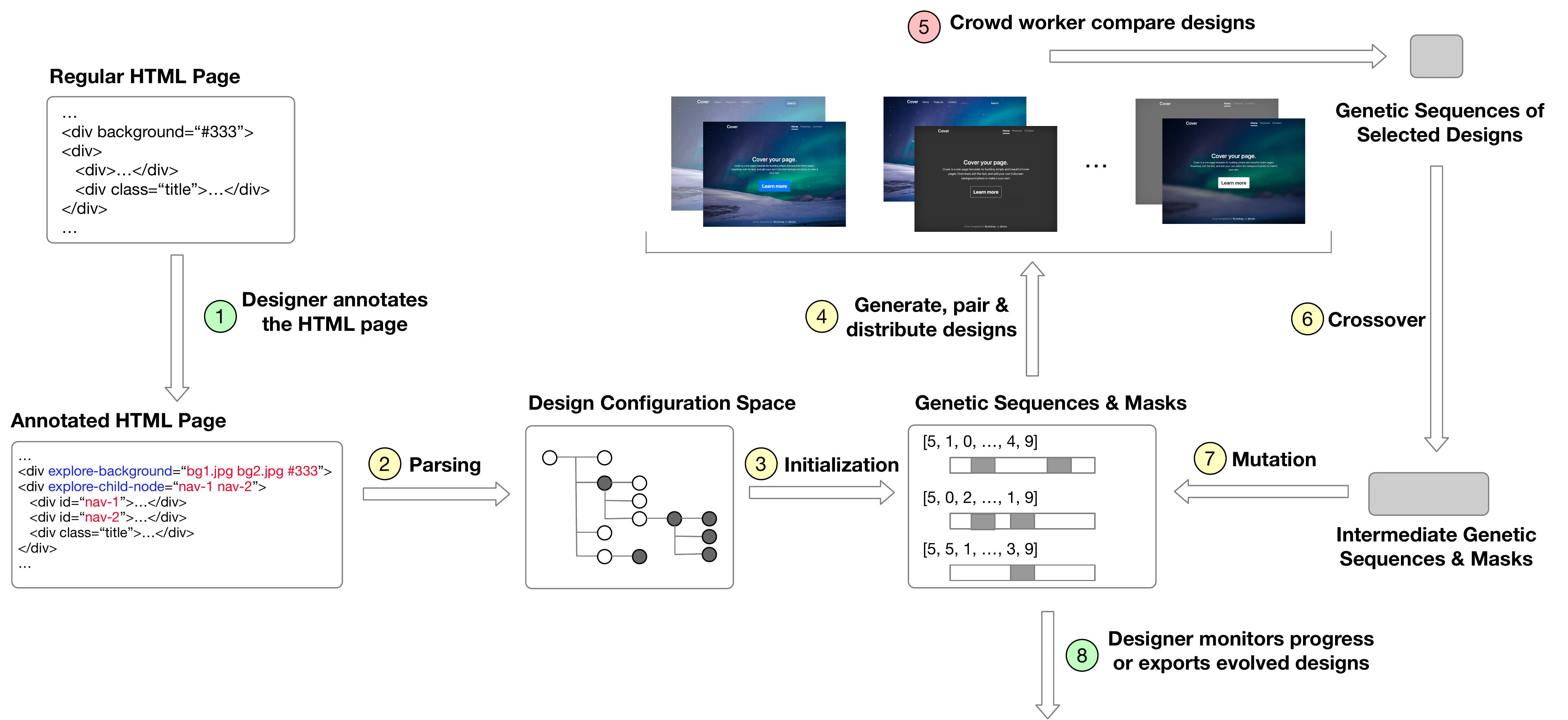}
  \caption{Spacewalker provides integrated support to allow a designer to rapidly explore a large design space of a webpage. Designer actions are \#1 and \#8 denoted in green, and worker actions are \#5 in red. The rest actions are performed by Spacewalker automatically. Spacewalker iterates through the loop of \#4-7 repeatedly to search the design space and evolve a design.}
  \Description[Spacewalker system diagram]{Step 1, designer annotates a regular HTML page. Step 2, Spacewalker parses the annotated HTML page into a design configuration space. Step 3, the enhanced genetic algorithm initializes, producing genetic sequences and masks. Step 4 through 7 forms a loop. In step 4, design examples are generated from the genetic sequences, and paired and distributed to crowd workers. In step 5, crowd workers compare designs and select one design for each pair. In step 6, the crossover operation is performed on the selected genetic sequences and masks. In step 7, the mutation operation is performed on the intermediate genetic sequences and masks, producing a new batch of genetic sequences and masks, which goes back to the loop. Meanwhile, designer monitors progress or exports evolved designs in step 8. }
  \label{fig:overview}
\end{figure*}

In this paper, we present Spacewalker, a tool that allows designers to rapidly search a design space of a web UI for an optimal design within that space (see Figure \ref{fig:overview}). In a typical HTML page or a CSS specification, designers first annotate each attribute they want to explore using a simple markup extension we designed. Our tool then parses the annotated HTML or CSS specification, and intelligently generates and distributes various configurations of the web UI to crowd workers for evaluation. Our research challenges are three-fold: 1) designing a markup annotation that is expressive and easy to use for specifying various design options, 2) developing an algorithm to allow efficient exploration of a large design space based on crowd worker feedback, and 3) creating a tool that can provide integrated support for design exploration.

To address challenge 1), We designed the HTML annotation as a simple extension of the existing HTML and CSS grammar, where instead of specifying a single value for an attribute, a designer can provide multiple candidate values for it, which are to be explored by Spacewalker. To address challenge 2), we enhanced a genetic algorithm by adding feedback mask-based stochastic sampling, to accommodate crowd worker responses from pairwise comparison of UI designs---that tends to yield more reliable feedback than rating each design separately. To address challenge 3), we created a web-based tool that streamlined the entire task of design exploration including task creation, monitoring and evaluation. 

We evaluated Spacewalker by asking interaction designers to use it for exploring a set of UI design tasks, and Spacewalker received positive feedback. To systematically examine how well Spacewalker algorithms can evolve a design by quickly searching a design space, we tested it on six design tasks that range in search space sizes and design types. The experiments indicate that UI designs obtained by Spacewalker were significantly more preferred by human evaluators\footnote{These evaluators were a separate group of crowd workers from those who were involved in human-in-the-loop design search.} than those from a baseline method. Our paper makes the following contributions:

\begin{itemize}
    \item An expressive and easy-to-use HTML markup extension that allows designers to easily specify various alternatives for design search, which requires negligible learning effort;
    \item An enhanced genetic algorithm that can efficiently explore a large design space using crowd worker responses from pairwise comparison of UI designs;
    \item Integrated general tool support that allows designers to easily obtain an improved design from a large range of options within a short period of time (e.g., 1 hour) at a small amount of cost (e.g., 35 US dollars).
\end{itemize}

\section{Related Work}

Our work is related to three areas of the literature, including UI evaluation methods, crowdsourcing-based design support, and interactive UI design optimization.

\subsection{Traditional UI Evaluation Methods}

Usability testing \cite{corry1997user} is a commonly used approach for evaluating a UI design, which often requires a user experience researcher to recruit user participants, moderate a study session, and observe and analyze findings from the study \cite{dumas1999practical}. 
To study how users react to design alternatives at scale, A/B testing is widely used where variants of a design are tested with different user populations and user behaviors are logged and statistically analyzed by user experience researchers \cite{Kohavi2017, kohavi2009controlled}. Various tools or platforms are available to support A/B testing of UI designs, such as GoodUI\footnote{https://goodui.org/}, Optimizely\footnote{https://www.optimizely.com/} and VWO\footnote{https://vwo.com/testing/ab-testing/}. 

Although existing methods are widely adopted, they often require substantial engineering effort to build and instrument a test. It also often involves extensive effort to analyze user data to extract findings that can be used for next design iteration. In addition, these methods are limited by the number of alternatives they can explore, which is problematic as a design space of a UI is often large. Consequently, an end design might be suboptimal due to limited exploration. 
AB4Web addresses this problem through randomized split testing, and successfully analyzed users' preferences for a task with 49 designs \cite{vanderdonckt2019ab4web}. Nevertheless, A/B testing still struggles to support a large design space when there are hundreds or thousands of design alternatives. In Spacewalker, we aim to address these issues by providing an integrated support for designers to explore a large design space and improve their design.


\subsection{Crowd-Powered UI Design \& Evaluation}

Previous work has incorporated crowdsourcing for UI design and evaluation. Crowdsourcing has shown success in providing comparable results for evaluating user interfaces with those acquired from a lab-based setting \cite{vliegendhart2012crowdsourced, komarov2013crowdsourcing}. Voyant allows designers to seek perception-oriented feedback from a non-expert crowd, with an emphasis on connecting the visual design with corresponding feedback \cite{xu2014voyant}. Reinecke et al. evaluated a set of 430 web designs through 40,000 online participants, demonstrating the feasibility of large-scale design evaluations through the crowd \cite{reinecke2014quantifying}. ZIPT allows designers to collect and visualize interaction patterns for any Android apps from the crowd \cite{deka2017zipt}. 

The crowd can be more actively involved in UI design tasks to provide feedback \cite{xu2014voyant, oppenlaender2020crowdui, wang2019kaleidoscope, reinecke2014quantifying} or participate in the design process \cite{lee2018exploring, park2013crowd, lee2017sketchexpress}. 
Apparition supports creating UI designs and animations from interface sketches and natural language descriptions through self-coordinated real-time crowdsourcing \cite{lasecki2015apparition}. Similar to previous work, we also embed the crowd in the loop of the design and evaluation process. However, we focus on the design task where an interaction designer has a basic HTML design and wants to obtain an optimal configuration for the design by exploring a large range of options such as colors, fonts and layouts. We also aim to minimize the effort and cost of the designer to perform the task.

\subsection{Interactive UI Design Optimization}
Using Artificial Intelligence algorithms to optimize interface design is a longstanding topic. Genetic algorithms (GA) in particular have been applied to optimize UI designs with large search spaces. Imagine generates style sheets for HTML pages interactively through user selection \cite{monmarche1999imagine}. Quiroz et al. combines GA with UI design metrics to reduce the number of choices needed by a user \cite{quiroz2007interactive}. However, these approaches only take the input from a few users, causing fatigue \cite{takagi2001interactive} and increasing potential bias. 

To address the issue, Salem \cite{salem2017user} combined crowdsourcing and genetic programming \cite{10.5555/522098} for the design of landing pages. Tamburrelli and Margara \cite{tamburrelli2014towards} explored approaches for optimizing software designs specified in Java through GA, basing their fitness function on the distance from users' interaction position. Despite the adoption of the crowd, these interactive GA solutions rely on implicit information, such as click location that is difficult to generalize to other design tasks. Moreover, the designer-facing tools require specific learning of a custom specification or programming language, which increases the burden on the designer. Although we employ GA-based algorithms and crowd in our work, Spacewalker is designed to address a general web UI design scenario. It allows designers to specify the design space for exploration using a simple extension of HTML tags. We also enhanced genetic programming for addressing worker responses from pairwise comparison of designs, which makes genetic programming more applicable for UI optimization.


\section{Using Spacewalker}
\label{sec:example}
We here describe how UI designers or developers would use Spacewalker to explore the design space of their user interfaces. Assume, Alex, an web designer is designing a new Product page for her company. Although she has written an HTML prototype of the page, she is uncertain about a few of design aspects of the page, such as color schemes, font sizes, and background choices. As the combination of these factors resembles a vast number of design alternatives, Alex decides to let Spacewalker to explore her design space.

To do so, Alex first edits the HTML prototype for the Product page by adding simple markup tags for the design options that she is unsure. For example, in the \verb/<div>/ element for the background, Alex adds the following tag:
\begin{alltt}
    <div \textbf{explore-background =
         "url(bg1.jpg) url(bg2.jpg) #333"}>
\end{alltt}
This instructs Spacewalker to explore three different alternatives for the page background: image "\verb/bg1.jpg/", image "\verb/bg2.jpg/", and a solid background with a dark gray color "\verb/#333/". Behind the scene, Spacewalker creates three page designs with each using a different background choice. Similarly, Alex can explore any CSS property for an HTML element, such as \verb/font-size/, \verb/margin/, \verb/height/ and \verb/width/, using a simple syntax rule, i.e., prefixing \textbf{explore-} to the property name.

In addition to exploring individual CSS properties, Alex wants to determine which design of the navigation bar she should use for the Product page. To do so, she adds an \verb/explore-child-id/ attribute to the parent \verb/div/ node of the navigation bar, which instructs Spacewalker to only use one out of multiple candidates when generating and evaluating a design.
\begin{alltt}
    <div \textbf{explore-child-id="nav-1 nav-2"}>
        <div \textbf{id="nav-1"}>    ... </div>
        <div \textbf{id="nav-2"}>    ... </div>
        <div class="title"> ... </div>
        ...
    </div>
\end{alltt}
In this example, Spacewalker uses either "\verb/nav-1/" or "\verb/nav-2/" at a time, while the rest children are unaffected. Note that the CSS attributes of elements for each node can be further explored, which enables recursive exploration.

Instead of specifying exploration strategies based on nodes, which can be tedious, designers can directly explore at the level of CSS specification using the \verb/explore-css/ tag. Here Alex would like to assure the titles (\verb/<h1>,<h2>/) and the body paragraphs (\verb/<p>/) are using matching colors that she designed, so she adds the options as a group:
\begin{alltt}
    <head> ...
        \textbf{<explore-css>}
            h1, h2: \{ color: (color1); \}
            p     : \{ color: (color2); \}
            --------
            h1, h2: \{ color: (color3); \}
            p     : \{ color: (color4); \}
        \textbf{</explore-css>}
    ...
    </head>
\end{alltt}
This ensures that Spacewalker would globally apply these options: either color1 for titles and color2 for body text, or color3 for titles and color4 for body text. A line of dashes (i.e., any number of "-") separates these two options.

\begin{figure}[h]
  \centering
  \fbox{\includegraphics[width=0.9\linewidth]{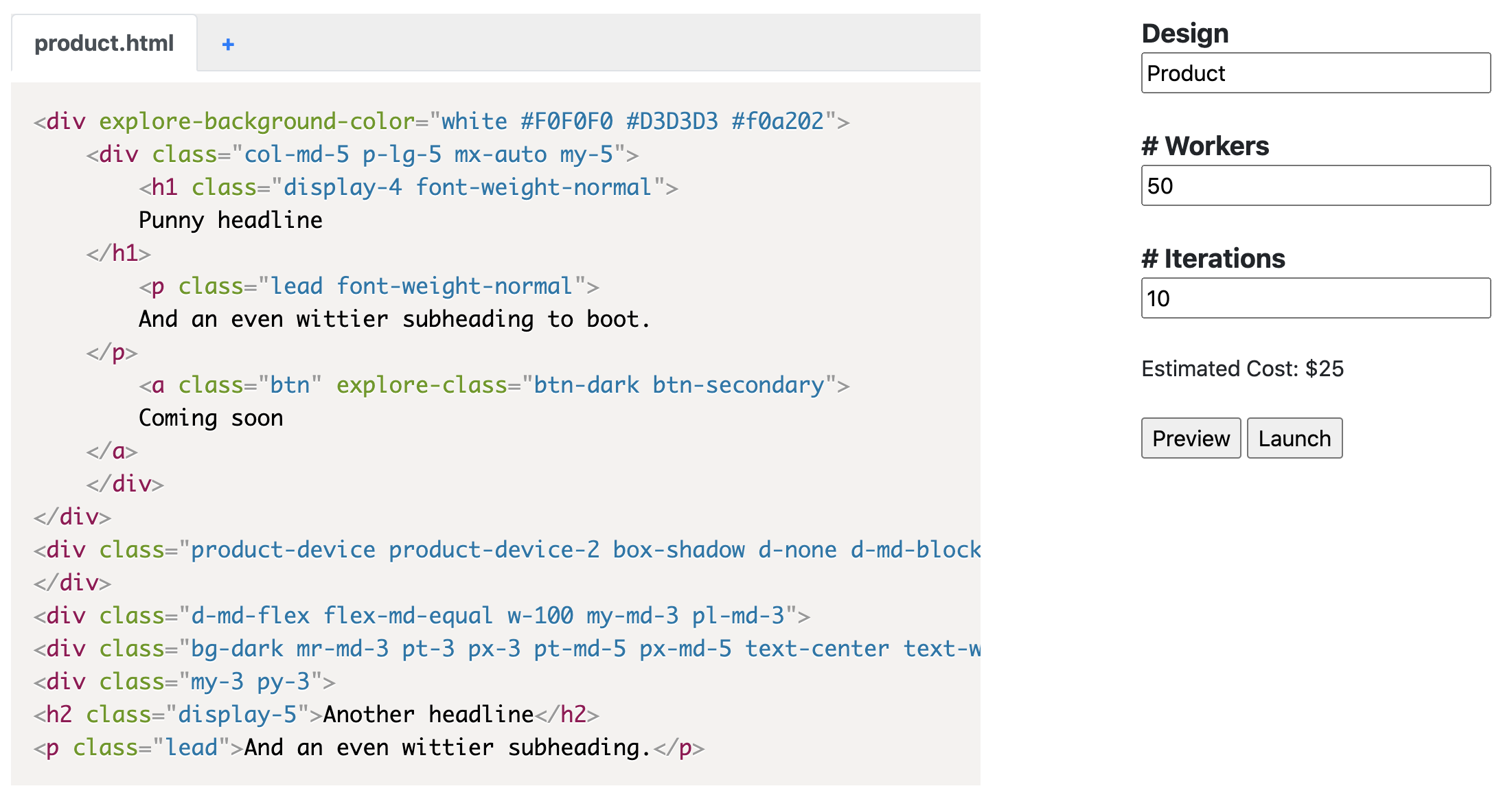}}
  \caption{The Author interface allows a designer to create and launch a task.}
  \Description{The Author interface. On the left, an annotated HTML code snippet is presented in a code editor. On the right, three text fields allow the user to specify the name of the design (Product in this case), the number of workers (50 in this case), and the number of iterations (10 in this case). The system provides an estimated cost, 25 dollars in this case. Two buttons are available: preview and launch.}
  \label{fig:author}
\end{figure}

\begin{figure}[h]
  \centering
  \fbox{\includegraphics[width=0.9\linewidth]{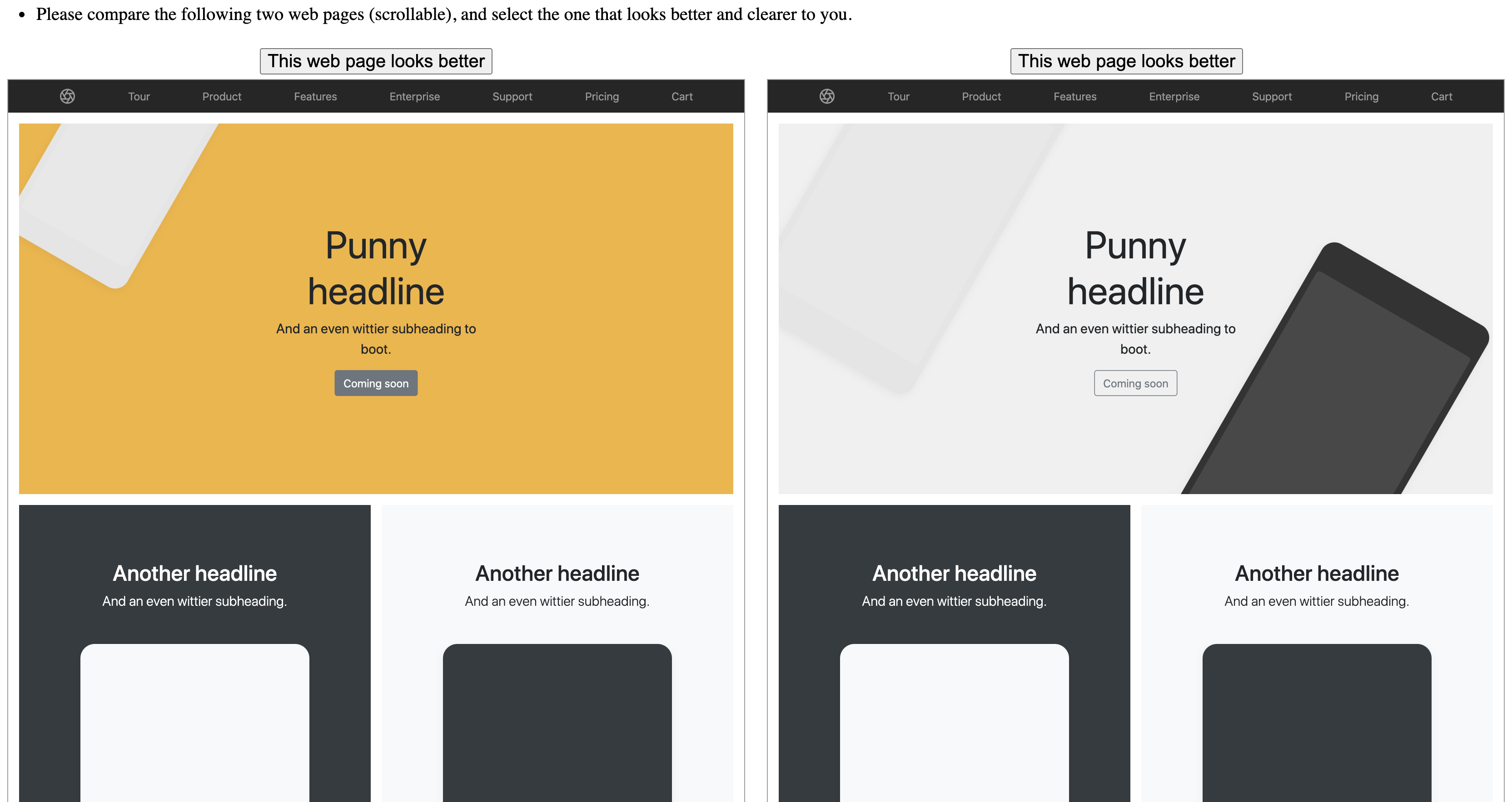}}
  \caption{The Eval interface allows a crowd worker to compare two design alternatives, and select the one they prefer.}
  \Description{The Eval interface. The prompt reads: "Please compare the following two web pages (scrollable), and select the one that looks better and clearer to you." Below are two web pages side by side, each with a button that reads: "this web page looks better."}
  \label{fig:worker}
\end{figure}

After creating the specifications for all the design aspects in questioning, Alex launches a Spacewalker task by specifying 50 human raters\footnote{In this paper, we use worker and rater interchangeably.} and 10 iterations (see Figure \ref{fig:author}). Spacewalker shows an estimate about how much it would cost for using the crowd workers. Alex can preview designs generated by Spacewalker based on her specifications. Upon Alex clicking the Launch button, Spacewalker automatically generates and sample designs based on her specification and distributes them to online raters that are recruited from Amazon MTurk. The interface for the rater is straightforward (see Figure \ref{fig:worker}). Raters see a pair of designs side by side, and are asked to select the one that they prefer.

\begin{figure}[h]
  \centering
  \fbox{\includegraphics[width=0.9\linewidth]{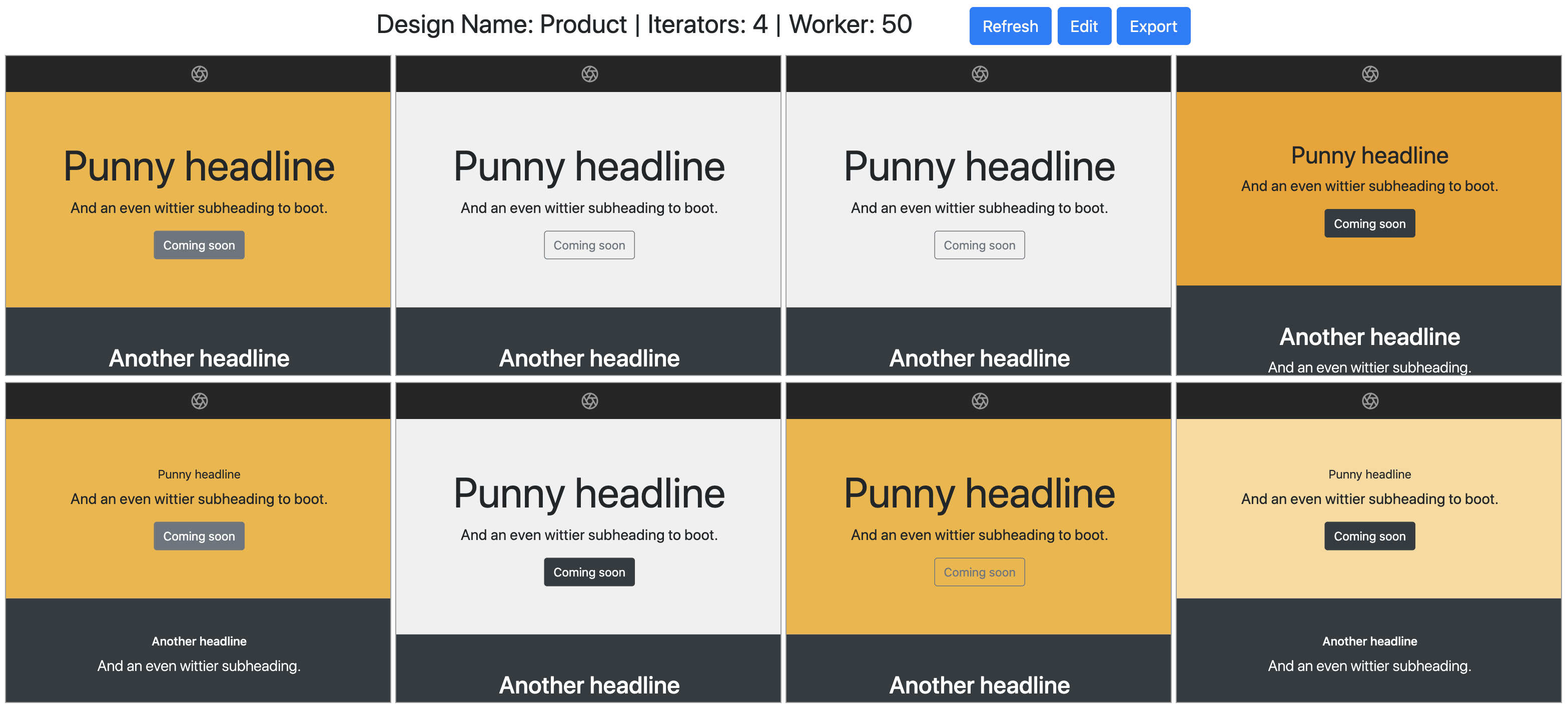}}
  \caption{The Progress viewer allows a designer to examine the progress of the task by viewing the designs generated from the current generation, and export the HTML specification of these designs if satisfied.}
  \Description{The Progress viewer. The heading reads: "Design name: product, iterations: 4, worker: 50". Three buttons are available on the right: refresh, edit, export. Below is a gallery view of eight designs from the current generation.}
  \label{fig:monitor}
\end{figure}

Alex can monitor the progress of the task in the Progress Viewer (see Figure \ref{fig:monitor}), which allows her to see the sample designs of the current generation (iteration). In about one hour, all the iterations are completed, and Alex selects five top designs from the collection of designs in the last generation. In case that none of the designs are satisfactory, Alex can edit the task in the Progress Viewer and relaunch the task to continue evolving the design. 

\section{The Spacewalker System}

In this section, we discuss the system design and algorithmic details that underline the Spacewalker. The Spacewalker system consists of three main components: an HTML specification parser, a genetic algorithm backend, and a crowdsourcing frontend. Once a designer submits an HTML specification, the parser extracts the attributes and options to be explored, which are passed on to the genetic algorithm. The genetic algorithm generates design instances from the options, which are sent to the crowdsourcing frontend to collect worker feedback. Once enough feedback is received for one iteration, the genetic algorithm generates the next generation of designs, and the process is repeated until the specified number of iterations is reached.

\subsection{Spacewalker Markup Syntax \& Parsing}
As shown in the above example, Spacewalker supports a rich set of methods for exploring a design space through simple HTML extensions, which are intuitive to designers as shown in our experiments. We here discuss its syntax and parsing details.

\subsubsection{Syntax}
\label{sec:syntax}
To explore a property of an individual element, a designer follows a simple syntax by prefixing "\verb/explore-/" to the property, and specifying the alternative values for the property delimited by spaces:
\begin{alltt}
    explore-\textbf{<property-name>} = 
      "\textbf{option-1 option-2 option-3} ..."
\end{alltt}

Spacewalker supports all CSS properties and any number of them for an element. If multiple properties need to be explored jointly (e.g., height and width), Spacewalker allows a designer to combine multiple properties for exploration by joining their names using "\verb/-and-/" and optional values using a semicolon (\verb/;/):
\begin{alltt}
    explore-\textbf{<property-A>}-and-\textbf{<property-B>} =
      "\textbf{option-A-1;option-B-1 option-A-2;option-B-2} ..."
\end{alltt}

In addition to explore individual elements, Spacewalker allows a designer to easily explore a large component of a design as a whole, which might contain a branch of elements and sub-trees, such as a side bar or a navigation bar. To do so, a designer can use the \verb/explore-child-id/ tag in a parent node with the id of each child that corresponds a design candidate as options. See Section~\ref{sec:example} for an example. Lastly, instead of exploring a design space based on elements, a designer can explore style options with CSS selectors using the same format as a regular CSS file, and by again prefixing the "\verb/explore-/" tag, and by using a line of dashes to indicate alternative styles (see examples in Section~\ref{sec:example}). Because of this, powerful CSS features, such as CSS variables (which can be used to store values in custom properties) \footnote{CSS variable on MDN Web Docs: \url{https://developer.mozilla.org/en-US/docs/Web/CSS/var()}}, can be adopted to streamline the specification of possible values for properties. For simplicity, we did not use these features in this paper.

\subsubsection{Parsing}
The Spacewalker parser analyzes a design specification file by parsing its HTML structure, which derives an internal representation for the design search space. It looks specifically for the \verb/explore-*/ tags, and records the options for each attribute provided by the designer. In addition, the parser adds a unique HTML \verb/id/ to elements without one, in order to link the attribute and options with the corresponding element. To preserve the hierarchical relationship of the HTML tree structure, the parser also maintains the hierarchical layout of the elements to be explored in a separate tree structure.

\subsection{Spacewalker Genetic Algorithms}

As the number of attributes and nodes to be explored increases, the search space for a design grows combinatorially. To search for an optimal design in the space, it is prohibitively expensive to examine every possible design configuration with worker evaluation. On the other hand, with a limited number of worker feedback, which is often the case in reality, a search very likely ends up with a sub-optimal design, as shown in our experiments later. As a result, it is necessary to use a more intelligent algorithm. We here focus on Genetic Algorithms, a popular choice that has been used in the literature, with several important enhancements. 


\subsubsection{Genetic Algorithm Background}

\begin{figure}[h]
  \centering
  \includegraphics[width=0.9\linewidth]{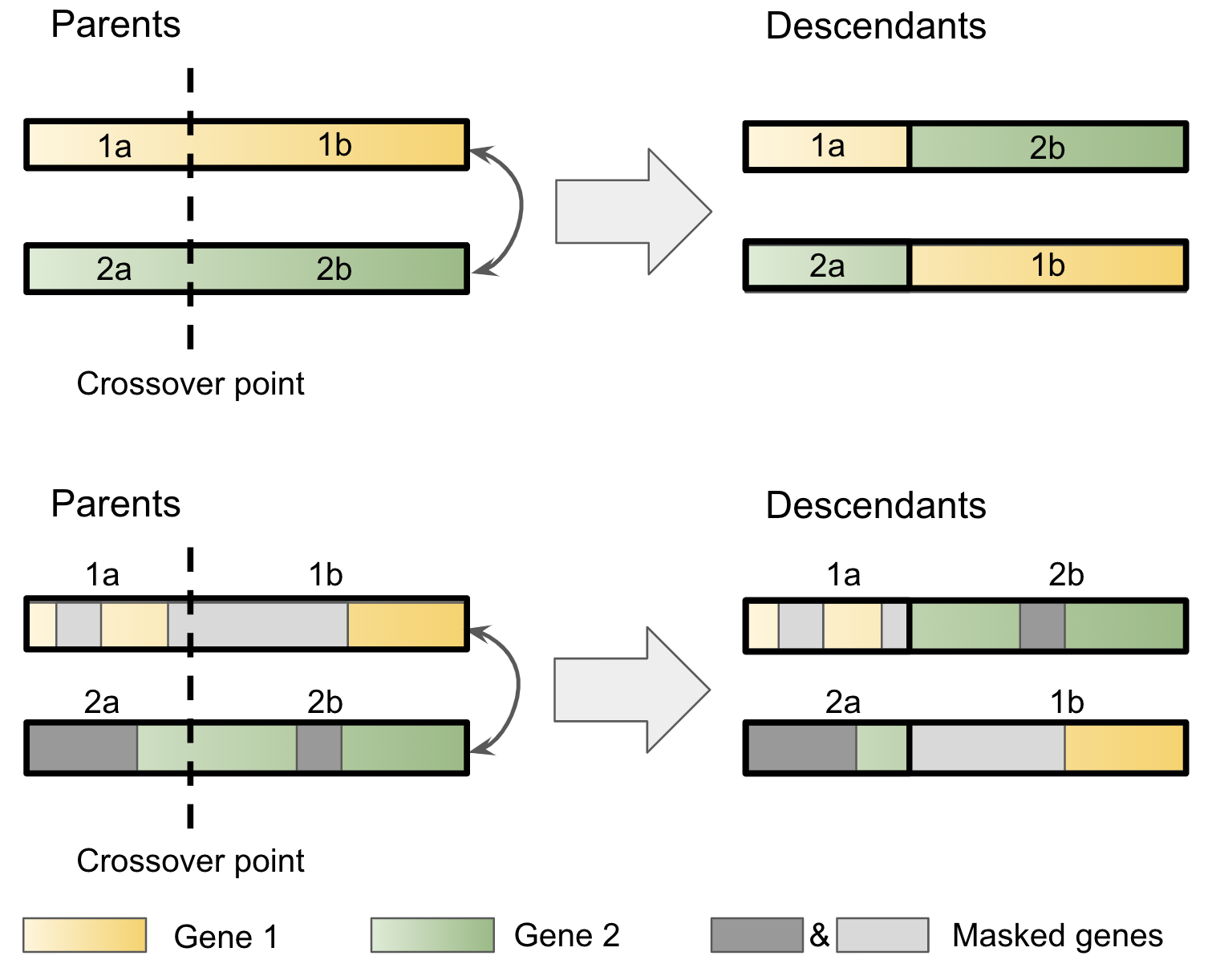}
  \caption{The crossover operation for traditional genetic algorithms (top) and for Spacewalker (bottom).}
  \Description{The crossover operation. The top illustrates traditional crossover operations. A crossover point is randomly selected from two parents, and all the genetic representation to the right of the crossover point are swapped, forming two children. The bottom illustrates the modified crossover operation in Spacewalker, where the masks for the parents are also swapped to the right of the crossover point.}
  \label{fig:ga}
\end{figure}

A typical genetic algorithm (GA) follows an iterative process, where potential solutions evolve from a multi-generation process. It consists of four stages: \textit{initialization}, \textit{selection}, \textit{crossover}, and \textit{mutation}. During \textit{initialization}, the first generation is randomly generated, with the goal of covering as many configurations as possible. Then, the algorithm loops through the rest of the stages, where each iteration leads to a new generation. During \textit{selection}, the algorithm selects a portion of the current population as the parents for the next generation based on a fitness function. After selecting the parents, the next generation is generated through the \textit{crossover} operation. Each time, a pair of parents are randomly selected by the fitness function, and their genetic representations are mixed based on a crossover operator. One method is single-point crossover, where a crossover point is randomly selected from both parents, and all the genetic representation to the right of the crossover point are swapped, forming two children (Figure~\ref{fig:ga}). Finally, the next generation goes through \textit{mutation}, where the genetic sequences are randomly altered to prevent the algorithm from running into a local minima.

\subsubsection{UI Design Search as Crowd-Driven Genetic Programming}
We refer an instance of a UI design, which is acquired by selecting a specific option for each attribute to be explored, as a \textit{configuration} of the design. Note that attributes that are not marked for exploration do not appear in a configuration for genetic programming because they are already determined by the designer. To adapt the genetic algorithm for searching an optimal UI configuration, we first encode each configuration as a genetic sequence, which is an ordered list of valued attributes whose value is denoted by the index to an option for the attribute. As an example, consider a specification that explores three attributes \verb/(A,B,C)/. If one configuration selected option \verb/3/ for attribute \verb/A/, option \verb/1/ for attribute \verb/B/, and option \verb/6/ for attribute \verb/C/, then the resulted genetic representation is \verb/[3,1,6]/. 

A specific genetic sequence indicates a UI configuration, which can be rendered as a design instance shown to a crowd worker for feedback. As a rater's judgement can be dominated by the early examples and may drift over time \cite{brochu2010bayesian}, it is generally difficult for a user to rate the goodness of a design with an absolute scale. Spacewalker instead presents each rater a pair of different candidate designs at a time, and asks the rater to select the preferred design, i.e., a two-alternative forced-choice (2AFC) method. Thus, our \textit{fitness function} outputs \verb/1/ for the preferred design and \verb/0/ for the less preferred one. Although presenting more than two examples in a gallery design can be another appealing alternative \cite{monmarche1999imagine, brochu2010bayesian}, the viewing area for each example would be too small in our case of web design, and may prevent raters from noticing  design details that matter.

\subsubsection{Enhancing Genetic Algorithms}
The pairwise comparison of designs eases the rater task and yields more reliable feedback than rating a design individually on an absolute scale. However, pairwise comparison raises challenges for genetic programming. The score that a design receives is now more with respect to the differences between two designs being compared, instead of every aspect of the design. In other words, the rater feedback carries information only for a subset of genes in a genetic sequence.

In our early exploration, we found conventional GA sensitive to the random initialization of design options. When the options shared by the pair of designs in comparison happen to be desirable, conventional GA yields good results. However, when these shared options are less desirable, conventional GA performs poorly as those options that are not compared also receive positive responses. To address the issue, we enhance traditional genetic algorithms, for each iteration, by directing rater feedback to genes that participate in a comparison while allowing the rest genes in a sequence to remain stochastic in the downstream evolution. 

To do so, we introduce a bit mask, named \textit{feedback mask}, for each genetic sequence---that corresponds a design instance, which has the same length as a genetic sequence. A bit in the feedback mask is \verb/1/ when the corresponding option is compared and favored by a crowd worker or \verb/0/ when it is not. For each pair of designs shown to a crowd worker, we compute a \textit{diff mask} to capture the differences between their genetic sequences where the corresponding bits for the differences are set to \verb/1/. The diff mask represents where we would likely gain knowledge by comparing these two designs: in the part where they differ. When the worker selects their preferred design, the original feedback mask for that design is combined with the diff mask through a bitwise logical \verb/OR/ operation. The resulting mask is assigned to the genetic sequence of the favored design, which captures both the information learned from previous generations (through the original feedback mask) and the current generation (through the diff mask). We elaborate on our algorithm below and particularly focus on how the masks operate in the initialization, crossover, and mutation phase.

\textbf{Initialization.}
The goal of our initialization process is to maximally cover possible options for each attribute, but meanwhile limit the total number of designs to be compared by human raters. Therefore, we treat each attribute independently during initialization, and generate design variations by sampling the options of one attribute at a time while affixing the rest exploratory attributes at a random value. The feedback mask for each genetic sequence is all zeros, as it has not received feedback from raters yet. Based on the result of each pairwise comparison, Spacewalker learns which design was preferred by a human rater to which the option contributes, and sets the corresponding mask value as \verb/1/, and the rest as \verb/0/.

\textbf{Crossover.}
We enhanced the single-point crossover operation of genetic algorithms, by assigning random option values to positions in the sequence (corresponding attributes in the design) where the feedback mask is \verb/0/ in the descendants. As a result, we introduce variations to a sequence where we have yet to acquire rater feedback. Meanwhile, in addition to crossover for the genetic sequences, the feedback masks also crossover at the sequence crossover point to carry the mask to the next generation.

\textbf{Mutation.}
Mutation is handled in a similar way to the traditional approach, where we alter one attribute in a genetic sequence based on a mutation rate. We chose a .03 mutation rate in Spacewalker, which is in line with other genetic algorithms applied on a similar population size. When an attribute is mutated, its mask is set to \verb/0/.

Finally, we consider nested designs (where one option value depend on another parent value). In this case, we link the child choices to the parent choices in the genes. When a parent is selected, only the relevant child genes would be active, and we only perform the crossover and mutation operations on these genes.

\subsection{Architectures}
The Spacewalker system is built as a web service based on AppEngine\footnote{\url{https://cloud.google.com/appengine}}. Our front end includes a task authoring interface for a designer to create and launch a task (see Figure~\ref{fig:author}), a monitor interface for the designer to monitor the task progress and export results (see Figure~\ref{fig:monitor}), and a worker interface for the worker to compare a pair of designs (see Figure~\ref{fig:worker}). Our backend parses a design specification, generates and distributes evaluation tasks, schedules workers and execute genetic operations on the sequences. A crowd worker first signs in the worker interface by entering their worker ID, and then performs a sequence of evaluation tasks in which each trial involves indicating their preference over a pair of designs. The back-end server is responsible for scheduling workers for different web UI pairs without conflicts and supporting multiple workers submitting results at the same time, using database read/write lock. When a worker is submitting a comparison result, or a new iteration is being generated by the genetic algorithm, the database is locked to ensure the atomicity of the operations. To the workers, the scheduling process is transparent and they always see a consistent labeling interface for comparing two web UIs.


\section{Experiments}
We evaluate Spacewalker in multiple dimensions. We conduct a user study to investigate whether Spacewalker markup extension to HTML is easy to understand and use by designers and developers, and how they react to the overall support of Spacewalker for design exploration. We also systematically examine how well Spacewalker explores a design space for designers and improve designs over iterations.

\subsection{User Study}
In this study, we evaluate the usability of our proposed HTML extensions by gather informal feedback from web designers. The goal is to test whether web designers are able to learn and use our markups to specify search criteria and launch a design exploration task, and to gather feedback of the Spacewalker system.

\subsubsection{Participants \& Procedure}
We invited five participants for this remote user study. Two of the participants were graduate students and the other three were professional developers. Four of them were trained in HCI and had experience with conducting user studies. All the participants indicated at least three years of experience with web interface design and development. These participants resemble HCI researchers and web developers who want to improve their design by quickly examining detailed design options with users at scale. 

We provided a description of supported functionalities and sample markup code snippets (similar to Section~\ref{sec:syntax}). We asked each participant to add markups to one template HTML web page (the \textit{Cover} example in Section~\ref{sec:exp_setup}), specifying exploration options for attributes or style sheet entries that they would like to change. We verbally walked them through the code snippets and demoed the usage, which took 10--15 minutes. Participants then edited the provided the HTML documents in their preferred code or text editors, and we recorded the time used for them to experiment with the markups and complete each task. After the study, each participant was asked to comment on their experience with learning the markups and creating the specification. 
We reviewed their completed HTML specifications to check if they were correct.

\subsubsection{Results \& Feedback}

\begin{table}
  \caption{The summary of User Study Results. The time includes participants both learning our markup extension, and creating their own specifications as well as inspecting the effects as adjusting them.}
  \label{tab:usability_result}
  \begin{tabular}{r|ccccc}
    \toprule
    Participant & P1 & P2 & P3 & P4 & P5 \\
    \midrule
    Number of attributes & 8 & 7 & 5 & 5 & 6 \\
    Search space size & 3888 & 2187 & 480 & 560 & 1152  \\
    Time (minutes) & 48 & 15 & 15 & 30 & 32 \\
  \bottomrule
\end{tabular}
\end{table}

All participants were able to learn the Spacewalker markup syntax using the description we provided and were able to create syntactically correct specifications. We analyzed the specifications submitted by the participants, and Table~\ref{tab:usability_result} summarizes the results. On average, participants were able to understand and create a search specification in 28 minutes (SD = 13.8), exploring different options for five to eight attributes. The search spaces specified ranged from 480 to 3888, which indicated the need of designers to explore large design spaces.

We received largely positive feedback from the participants. P1 and P4 reported that the syntax was easy to learn, "even with basic knowledge background about HTML and CSS" (P4). P1 praised the system for "supporting all existing CSS properties". All participants appreciated the time savings and improved efficiency when working with Spacewalker. In particular, P1, P4, and P5 found Spacewalker to require less effort than the the traditional way of exploring multiple designs individually.


\subsection{Exploration Performance Evaluation}

We conducted two experiments to evaluate whether Spacewalker was able to efficiently search a design space and generate better designs by utilizing the responses from the crowd workers. We compare Spacewalker genetic algorithm against a baseline method that uniformly samples the design space for crowd evaluation. In the first experiment, we examine the effect of different search space sizes on the techniques. In the second experiment, we test these search methods on different types of web pages.

\subsubsection{Experimental Procedures \& Setup}
\label{sec:exp_setup}

We conducted both experiments following the same procedure. 
For Spacewalker, we used 10 iterations with 50 design samples in each iteration, which requires 25 comparisons by the raters. To reduce the potential influence from a single worker, we used workers who had above 90\% approval rate in MTurk, and limited each worker to performing 5 comparisons (10 samples). Therefore, each design search task needed 50 raters. To account for raters that may not be responsive after accepting the tasks, we distributed the tasks to 70 raters. For uniform sampling, to ensure that it receives the same number of rater responses as the genetic method, we randomly deployed 500 samples, which amounts to 250 pairs thus 250 rater responses. We also used 70 raters here to ensure enough responses. In sum, the only difference between the two conditions is the method used for searching the design space, while the rest aspects including the feedback mechanism is the same. We compensated each rater that finished the 5 comparisons 0.5 US dollars. A rater was only allowed to work in each method condition once.

On average, each task took about 1 hour to finish. After a task was finished, we selected the five designs that received most votes from raters for each method. For Spacewalker, these designs were drawn from the population of the last generation. For uniform sampling, they were chosen by ranking all the selected samples that were returned. We then deployed another task to a separate group of crowd workers for evaluating the quality of these selected designs. Each rater in the evaluation was asked to compare designs from Spacewalker genetic method and those from uniform sampling side by side. We refer to this round of crowd tasks as the \textit{cross-method evaluation}. The presentation order was randomized, and the rater had no knowledge of the underlying method of each design. We also randomly shuffled the order of designs for both methods. We ran 100 comparison tasks for each search specification, and the rater must make a choice between one of the two designs. The setup of the two experiments is the following.

\noindent \textbf{Experiment \#1: Effects of Search Space Sizes.}
In this experiment, we varied the search space size by using a different number of attributes and options in the design search specification. We based our study on the \textit{Cover} example provided by Bootstrap\Bootstrap, and we added Spacewalker markups to create the specifications used in the study. The number of options being explored ranged from 3 to 8 in this example, which corresponds a search space size ranging from 50 to 11,000 (see Table~\ref{tab:exp1_result} for all the search space sizes).

\noindent \textbf{Experiment \#2: Effects of Web Page Designs.}
We added exploration attributes and options to five additional web page templates, which are also based on Bootstrap examples\Bootstrap. The specifications for these designs were created so that their search spaces are similar in size (between 972 and 1215, mean=1050) for all the tasks. We used the following templates in this experiment (search space sizes in parentheses): \textit{Album} (972), \textit{Blog} (1080), \textit{Cover} (972), \textit{Dashboard} (1215), Pricing (1056), Product (1008). 


\subsubsection{Performance Results}

\begin{table*}
  \caption{Rater Preference for Spacewalker in Experiment 1}
  \label{tab:exp1_result}
  \begin{tabular}{c|cccccc}
    \toprule
    Search space size & 50 & 200 & 500 & 1000 & 3000 & 11,000\\
    \midrule
    Percentage of votes (\%) & 66 & 60 & 73 & 78 & 80 & 75 \\
    Z-score & 3.36 & 2.55 & 4.77 & 6.72 & 7.20 & 5.74 \\
    $p$ (two-tailed) & <.001 & .01 & < .001 & < .001 & < .001 & < .001 \\
  \bottomrule
\end{tabular}
\end{table*}

\begin{table*}
  \caption{Rater Preference for Spacewalker in Experiment 2}
  \label{tab:exp2_result}
  \begin{tabular}{c|cccccc}
    \toprule
    Page name & \textit{Album} & \textit{Blog} & \textit{Cover} & \textit{Dashboard} & \textit{Pricing} & \textit{Product} \\
    \midrule
    Percentage of votes (\%) & 71 & 75 & 78 & 64 & 59 & 70 \\
    Z-score & 4.60 & 5.74 & 6.72 & 2.90 & 2.00 & 4.24 \\
    $p$ (two-tailed) & <.001 & <.001 & <.001 & .005 & .045 & <.001 \\
  \bottomrule
\end{tabular}
\end{table*}

\begin{figure}[h]
  \centering
  \includegraphics[width=0.9\linewidth]{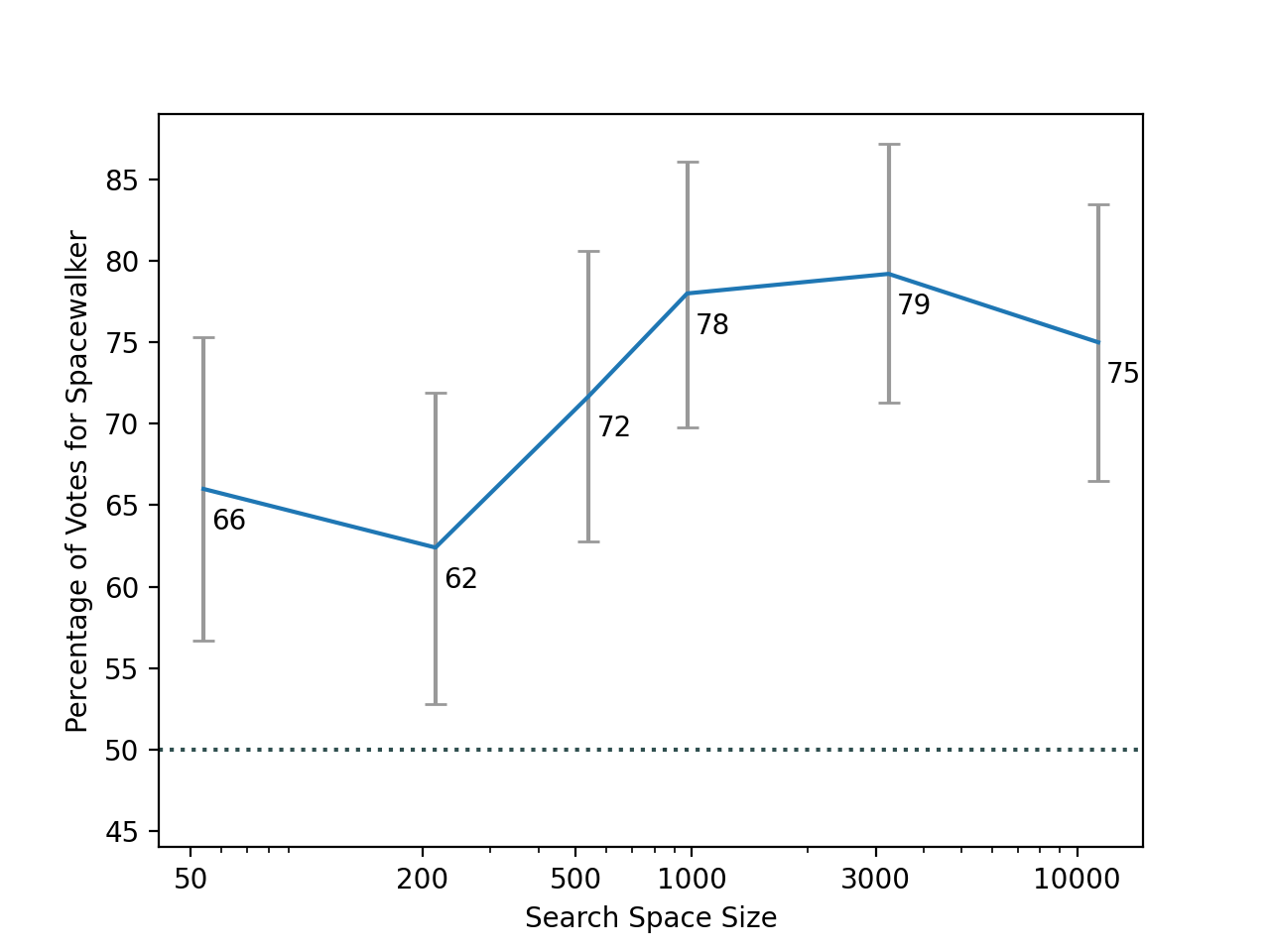}
  \caption{Voter preference for Spacewalker with varying search space sizes. The horizontal axis uses a log scale.}
  \Description{Percentage of votes for spacewalker is 66 for search space size 50, 62 for search space size 200, 72 for search space size 500, 78 for search space size 1000, 79 for search space size 3000, and 75 for search space size 10000. The overall trend is going up.}
  \label{fig:search_space}
\end{figure}

For each search specification, we calculate the percentage of votes received by Spacewalker genetic method (the rest of the votes are received by uniform sampling). We also conduct one-sample z-tests on the differences between a random draw and the voter's preferences for Spacewalker in each search specification.

Crowd raters, from the cross-method evaluation, showed significant preference for the designs generated by Spacewalker for all the search space sizes in Experiment~1 (see Table~\ref{tab:exp1_result}). In addition, we observe a trend of increased preference for Spacewalker genetic method as the size of the search space increases 
(see Figure~\ref{fig:search_space}). This indicates that the larger the search space is the more benefit there is for using Spacewalker.

\begin{figure*}[!htbp]
  \centering
  \includegraphics[width=0.976\linewidth]{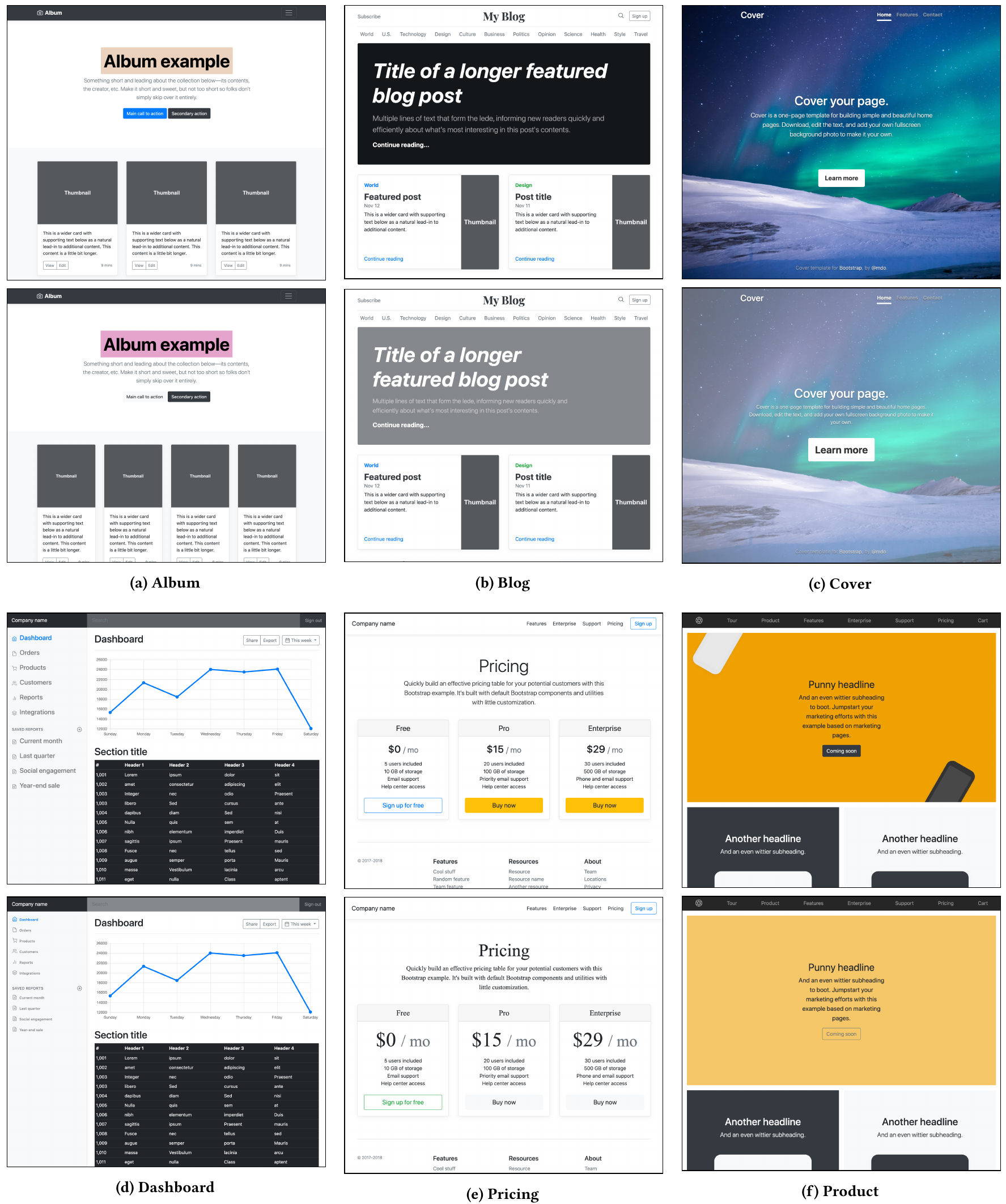}
  \caption{The top designs generated by Spacewalker (top) versus those from uniform sampling (bottom) for each web page template. These pages were adapted from the Bootstrap examples (see Section~\ref{sec:exp_setup}).}
  \Description{Top design examples for album, blog, cover, dashboard, pricing, product, respectively, comparing between Spacewalker and uniform sampling.}
\label{fig:examples}
\end{figure*}

For the experiment where Spacewalker is used to search for different web page types, we find that crowd raters, from the cross-method evaluation, preferred the results produced by Spacewalker genetic method in all cases we tested when they are compared with those from the uniform sampling method (see Table~\ref{tab:exp2_result}). Figure~\ref{fig:examples} shows the top designs generated by Spacewalker and those from uniform sampling for each of the web pages. Note that depending on the search options specified in a design, the difference between the outcome designs from the two methods can be subtle sometime, e.g., the Dashboard case in Figure~\ref{fig:examples}. Nevertheless, there is still strong consistency in raters preferences towards Spacewalker genetic method. This indicates that the benefit of Spacewalker is well demonstrated across different web page types.

\section{Discussion \& Future Work}
In this section, we discuss the strengths and limitation of our work, and our plan for future work. Our experiments show that the concept of Spacewalker is well received by the designers and developers. They feel Spacewalker is highly valuable for the design task. As P4 commented:

\textit{"This tool provides (a) useful way to compare my designs. I used to use the Inspect tool in Chrome to try out different values of the styles of my attributes, but the limitation is that I can only modify one item at a time. With this tool I could manage my HTML/CSS code and potential designs of the whole page efficiently. It improves my productivity and experience significantly."}

The Spacewalker markup extension is easy to understand and use for specifying design exploration. Our participants gave us several useful suggestions for improvements. P1 suggested that Spacewalker should provide suggestions for possible options to explore for a specific property, and warns designers when an option value is out of a reasonable range for a good design. This will require Spacewalker to encode certain design knowledge to make proper suggestions. Designers also want to immediately see the effect of a design when adding an option value, instead of examining them on a separate screen in the Preview (see Figure~\ref{fig:author}).

Another challenge lies in how well designers can understand the effect when complex design alternatives exist in one design space, e.g., design options nested within a parent option or global options via CSS. Although our participants did not encounter much difficulty, this can be challenging as the design space becomes convoluted. We believe the above extensions can provide a good starting point for designers to understand the search space. In addition, designers would be able to easily include or exclude certain combinations given appropriate visualization tools, which can be utilized by Spacewalker as an initialization condition.

Dependency between elements and designer specified options also presents two challenges to Spacewalker. First, in complex design spaces, designers may want to maintain dependencies between several sets of elements, while style options for different elements can also be dependent, where only certain options or elements can be combined together. In order to support more advanced dependency, we believe automatic tooling for identifying option dependencies and detecting potential inconsistency would be necessary, which provides opportunity for future work. Second, with multiple dependencies and the resulting hierarchical design spaces, the search space for a design grows combinatorially. Therefore, even our enhanced genetic algorithm may not converge to an optimal solution with a small number of comparisons. The gist is how to search the vast space efficiently under the monetary and time constraint for a task. However, we note that Spacewalker performed well on rather large search spaces in our experiment, and that other approaches, such as uniform sampling, would suffer more in these cases as the probability of encountering one "good" example would be minuscule. With a large enough search space, the effectiveness of any algorithm will be impacted. Effectively conveying such expectation to users is essential. Moreover, the system can offer to break down large multi-level search spaces into smaller ones, and perform design search tasks on each of them.

Our quantitative experiments for examining the performance of Spacewalker algorithms for searching a design space indicate that it improves designs over time by producing better design candidates,  particularly when the search space is large. However, we observed that the quality of these designs could further improved with more iterations and workers. These improvements can be easily achievable, particularly because these were achieved by only using 50 workers and a small monetary budget (around 35 US Dollars), which gives a lot of room to scale up. 

For each of the tasks, our GA-based algorithm only visited a small portion of the design space. 250 comparisons were made for each design. The search space size of our designs ranges from 50 to 10000. Combinatorially, the number of pairwise comparisons needed to cover the smallest search space (50) is $C_{50}^{2}=1,225$. Although searching over the entire space would be ideal, it is impractical to do so. Our search algorithm searches efficiently given a budget that dictates the number of comparisons and raters can be used. 

In addition, there is a research opportunity to better support designers by providing more guidance when creating a design search task. In Spacewalker, a designer currently needs to specify the number of iterations and workers needed for searching a design space (see Figure~\ref{fig:author}). A designer can specify them based on the complexity of their design task, and their time and budget constraints. However, it might not be always clear to the designer what the reasonable values are for these parameters. Currently, a designer can revise these parameters by examining the resulted designs shown in the Monitor view (see Figure~\ref{fig:monitor}) and schedule more iterations and workers to the task to improve the designs. We plan to infer these hyperparamters (the number of workers and iterations) needed automatically. For example, a general rule of thumb for the genetic population size for each generation (iteration) is 10 times the dimensionality of the search space \cite{storn1996usage}. We can train a model to predict a reasonable worker and iteration size by incorporating design complexities. These are beyond the scope of this paper.  


Finally, we want to emphasize that Spacewalker is able to identify an optimal design in a defined design space, instead of finding a globally optimal design. Spacewalker is not intended as a tool for creating a design from scratch. The target use case of Spacewalker is when a designer already has a basic design, and wants to examine a set of design alternatives or dimensions in mind, e.g., colors, styles and layouts, in light of user feedback. Conceptually, Spacewalker plays a similar role as A/B testing. However, it tremendously enhances A/B testing by allowing designers to explore a much larger design space with minimal effort. It allows designers to easily instrument a design exploration using our simple markup extension, seamlessly distribute design critique tasks to crowd workers, and quickly receive exploration results with genetic algorithms. The streamlined end-to-end support of Spacewalker is useful for both novice and experienced designers. To further our understanding, we intend to compare the automatic process enabled by Spacewalker with the existing manual process for design search in realistic design projects in the future.

\section{Conclusion}

Spacewalker provides integrated support to enable designers to rapidly explore a large design space to improve their web UI design. Our HTML markup for creating exploration specification provides a lightweight and familiar language for designers to specify complex designs and search requirements. By adapting genetic algorithms to effectively utilize crowd worker feedback, our system can quickly explore the search space of a web design, which provides real-time feedback to the designer about the progress of the search. Our experiments indicate that Spacewalker is well received by designers, and Spacewalker's genetic search algorithm significantly outperformed a uniform sampling baseline under different search space sizes and web design types.

\section*{Acknowledgements}
We would like to thank anonymous reviewers for their insightful feedback for improving the paper. We would also like to thank the participants in our user studies.

\bibliographystyle{ACM-Reference-Format}
\bibliography{main}


\begin{thebibliography}{26}


\ifx \showCODEN    \undefined \def \showCODEN     #1{\unskip}     \fi
\ifx \showDOI      \undefined \def \showDOI       #1{#1}\fi
\ifx \showISBNx    \undefined \def \showISBNx     #1{\unskip}     \fi
\ifx \showISBNxiii \undefined \def \showISBNxiii  #1{\unskip}     \fi
\ifx \showISSN     \undefined \def \showISSN      #1{\unskip}     \fi
\ifx \showLCCN     \undefined \def \showLCCN      #1{\unskip}     \fi
\ifx \shownote     \undefined \def \shownote      #1{#1}          \fi
\ifx \showarticletitle \undefined \def \showarticletitle #1{#1}   \fi
\ifx \showURL      \undefined \def \showURL       {\relax}        \fi
\providecommand\bibfield[2]{#2}
\providecommand\bibinfo[2]{#2}
\providecommand\natexlab[1]{#1}
\providecommand\showeprint[2][]{arXiv:#2}

\bibitem[\protect\citeauthoryear{Brochu, Brochu, and de~Freitas}{Brochu
  et~al\mbox{.}}{2010}]%
        {brochu2010bayesian}
\bibfield{author}{\bibinfo{person}{Eric Brochu}, \bibinfo{person}{Tyson
  Brochu}, {and} \bibinfo{person}{Nando de Freitas}.}
  \bibinfo{year}{2010}\natexlab{}.
\newblock \showarticletitle{A Bayesian Interactive Optimization Approach to
  Procedural Animation Design}. In \bibinfo{booktitle}{\emph{Proceedings of the
  2010 ACM SIGGRAPH/Eurographics Symposium on Computer Animation}} (Madrid,
  Spain) \emph{(\bibinfo{series}{SCA '10})}. \bibinfo{publisher}{Eurographics
  Association}, \bibinfo{address}{Goslar, DEU}, \bibinfo{pages}{103–112}.
\newblock


\bibitem[\protect\citeauthoryear{Corry, Frick, and Hansen}{Corry
  et~al\mbox{.}}{1997}]%
        {corry1997user}
\bibfield{author}{\bibinfo{person}{Michael~D Corry},
  \bibinfo{person}{Theodore~W Frick}, {and} \bibinfo{person}{Lisa Hansen}.}
  \bibinfo{year}{1997}\natexlab{}.
\newblock \showarticletitle{User-centered design and usability testing of a web
  site: An illustrative case study}.
\newblock \bibinfo{journal}{\emph{Educational technology research and
  development}} \bibinfo{volume}{45}, \bibinfo{number}{4}
  (\bibinfo{year}{1997}), \bibinfo{pages}{65--76}.
\newblock


\bibitem[\protect\citeauthoryear{Deka, Huang, Franzen, Nichols, Li, and
  Kumar}{Deka et~al\mbox{.}}{2017}]%
        {deka2017zipt}
\bibfield{author}{\bibinfo{person}{Biplab Deka}, \bibinfo{person}{Zifeng
  Huang}, \bibinfo{person}{Chad Franzen}, \bibinfo{person}{Jeffrey Nichols},
  \bibinfo{person}{Yang Li}, {and} \bibinfo{person}{Ranjitha Kumar}.}
  \bibinfo{year}{2017}\natexlab{}.
\newblock \showarticletitle{ZIPT: Zero-Integration Performance Testing of
  Mobile App Designs}. In \bibinfo{booktitle}{\emph{Proceedings of the 30th
  Annual ACM Symposium on User Interface Software and Technology}} (Qu\'{e}bec
  City, QC, Canada) \emph{(\bibinfo{series}{UIST '17})}.
  \bibinfo{publisher}{Association for Computing Machinery},
  \bibinfo{address}{New York, NY, USA}, \bibinfo{pages}{727–736}.
\newblock
\showISBNx{9781450349819}
\urldef\tempurl%
\url{https://doi.org/10.1145/3126594.3126647}
\showDOI{\tempurl}


\bibitem[\protect\citeauthoryear{Dumas and Redish}{Dumas and Redish}{1999}]%
        {dumas1999practical}
\bibfield{author}{\bibinfo{person}{Joseph~S. Dumas} {and}
  \bibinfo{person}{Janice~C. Redish}.} \bibinfo{year}{1999}\natexlab{}.
\newblock \bibinfo{booktitle}{\emph{A Practical Guide to Usability Testing}
  (\bibinfo{edition}{1st} ed.)}.
\newblock \bibinfo{publisher}{Intellect Books}, \bibinfo{address}{GBR}.
\newblock
\showISBNx{1841500208}


\bibitem[\protect\citeauthoryear{Iitsuka and Matsuo}{Iitsuka and
  Matsuo}{2015}]%
        {iitsuka2015website}
\bibfield{author}{\bibinfo{person}{Shuhei Iitsuka} {and}
  \bibinfo{person}{Yutaka Matsuo}.} \bibinfo{year}{2015}\natexlab{}.
\newblock \showarticletitle{Website Optimization Problem and Its Solutions}. In
  \bibinfo{booktitle}{\emph{Proceedings of the 21th ACM SIGKDD International
  Conference on Knowledge Discovery and Data Mining}} (Sydney, NSW, Australia)
  \emph{(\bibinfo{series}{KDD '15})}. \bibinfo{publisher}{Association for
  Computing Machinery}, \bibinfo{address}{New York, NY, USA},
  \bibinfo{pages}{447–456}.
\newblock
\showISBNx{9781450336642}
\urldef\tempurl%
\url{https://doi.org/10.1145/2783258.2783351}
\showDOI{\tempurl}


\bibitem[\protect\citeauthoryear{Kohavi and Longbotham}{Kohavi and
  Longbotham}{2017}]%
        {Kohavi2017}
\bibfield{author}{\bibinfo{person}{Ron Kohavi} {and} \bibinfo{person}{Roger
  Longbotham}.} \bibinfo{year}{2017}\natexlab{}.
\newblock \bibinfo{booktitle}{\emph{Online Controlled Experiments and A/B
  Testing}}.
\newblock \bibinfo{publisher}{Springer US}, \bibinfo{address}{Boston, MA},
  \bibinfo{pages}{922--929}.
\newblock
\showISBNx{978-1-4899-7687-1}
\urldef\tempurl%
\url{https://doi.org/10.1007/978-1-4899-7687-1_891}
\showDOI{\tempurl}


\bibitem[\protect\citeauthoryear{Kohavi, Longbotham, Sommerfield, and
  Henne}{Kohavi et~al\mbox{.}}{2009}]%
        {kohavi2009controlled}
\bibfield{author}{\bibinfo{person}{Ron Kohavi}, \bibinfo{person}{Roger
  Longbotham}, \bibinfo{person}{Dan Sommerfield}, {and}
  \bibinfo{person}{Randal~M Henne}.} \bibinfo{year}{2009}\natexlab{}.
\newblock \showarticletitle{Controlled experiments on the web: survey and
  practical guide}.
\newblock \bibinfo{journal}{\emph{Data mining and knowledge discovery}}
  \bibinfo{volume}{18}, \bibinfo{number}{1} (\bibinfo{year}{2009}),
  \bibinfo{pages}{140--181}.
\newblock


\bibitem[\protect\citeauthoryear{Komarov, Reinecke, and Gajos}{Komarov
  et~al\mbox{.}}{2013}]%
        {komarov2013crowdsourcing}
\bibfield{author}{\bibinfo{person}{Steven Komarov}, \bibinfo{person}{Katharina
  Reinecke}, {and} \bibinfo{person}{Krzysztof~Z. Gajos}.}
  \bibinfo{year}{2013}\natexlab{}.
\newblock \showarticletitle{Crowdsourcing Performance Evaluations of User
  Interfaces}. In \bibinfo{booktitle}{\emph{Proceedings of the SIGCHI
  Conference on Human Factors in Computing Systems}} (Paris, France)
  \emph{(\bibinfo{series}{CHI '13})}. \bibinfo{publisher}{Association for
  Computing Machinery}, \bibinfo{address}{New York, NY, USA},
  \bibinfo{pages}{207–216}.
\newblock
\showISBNx{9781450318990}
\urldef\tempurl%
\url{https://doi.org/10.1145/2470654.2470684}
\showDOI{\tempurl}


\bibitem[\protect\citeauthoryear{Lasecki, Kim, Rafter, Sen, Bigham, and
  Bernstein}{Lasecki et~al\mbox{.}}{2015}]%
        {lasecki2015apparition}
\bibfield{author}{\bibinfo{person}{Walter~S. Lasecki}, \bibinfo{person}{Juho
  Kim}, \bibinfo{person}{Nick Rafter}, \bibinfo{person}{Onkur Sen},
  \bibinfo{person}{Jeffrey~P. Bigham}, {and} \bibinfo{person}{Michael~S.
  Bernstein}.} \bibinfo{year}{2015}\natexlab{}.
\newblock \showarticletitle{Apparition: Crowdsourced User Interfaces That Come
  to Life as You Sketch Them}. In \bibinfo{booktitle}{\emph{Proceedings of the
  33rd Annual ACM Conference on Human Factors in Computing Systems}} (Seoul,
  Republic of Korea) \emph{(\bibinfo{series}{CHI '15})}.
  \bibinfo{publisher}{Association for Computing Machinery},
  \bibinfo{address}{New York, NY, USA}, \bibinfo{pages}{1925–1934}.
\newblock
\showISBNx{9781450331456}
\urldef\tempurl%
\url{https://doi.org/10.1145/2702123.2702565}
\showDOI{\tempurl}


\bibitem[\protect\citeauthoryear{Lee, Krosnick, Park, Keelean, Vaidya, O'Keefe,
  and Lasecki}{Lee et~al\mbox{.}}{2018}]%
        {lee2018exploring}
\bibfield{author}{\bibinfo{person}{Sang~Won Lee}, \bibinfo{person}{Rebecca
  Krosnick}, \bibinfo{person}{Sun~Young Park}, \bibinfo{person}{Brandon
  Keelean}, \bibinfo{person}{Sach Vaidya}, \bibinfo{person}{Stephanie~D
  O'Keefe}, {and} \bibinfo{person}{Walter~S Lasecki}.}
  \bibinfo{year}{2018}\natexlab{}.
\newblock \showarticletitle{Exploring real-time collaboration in crowd-powered
  systems through a ui design tool}.
\newblock \bibinfo{journal}{\emph{Proceedings of the ACM on Human-Computer
  Interaction}} \bibinfo{volume}{2}, \bibinfo{number}{CSCW}
  (\bibinfo{year}{2018}), \bibinfo{pages}{1--23}.
\newblock


\bibitem[\protect\citeauthoryear{Lee, Zhang, Wong, Yang, O'Keefe, and
  Lasecki}{Lee et~al\mbox{.}}{2017}]%
        {lee2017sketchexpress}
\bibfield{author}{\bibinfo{person}{Sang~Won Lee}, \bibinfo{person}{Yujin
  Zhang}, \bibinfo{person}{Isabelle Wong}, \bibinfo{person}{Yiwei Yang},
  \bibinfo{person}{Stephanie~D. O'Keefe}, {and} \bibinfo{person}{Walter~S.
  Lasecki}.} \bibinfo{year}{2017}\natexlab{}.
\newblock \showarticletitle{SketchExpress: Remixing Animations for More
  Effective Crowd-Powered Prototyping of Interactive Interfaces}. In
  \bibinfo{booktitle}{\emph{Proceedings of the 30th Annual ACM Symposium on
  User Interface Software and Technology}} (Qu\'{e}bec City, QC, Canada)
  \emph{(\bibinfo{series}{UIST '17})}. \bibinfo{publisher}{Association for
  Computing Machinery}, \bibinfo{address}{New York, NY, USA},
  \bibinfo{pages}{817–828}.
\newblock
\showISBNx{9781450349819}
\urldef\tempurl%
\url{https://doi.org/10.1145/3126594.3126595}
\showDOI{\tempurl}


\bibitem[\protect\citeauthoryear{Mitchell}{Mitchell}{1998}]%
        {10.5555/522098}
\bibfield{author}{\bibinfo{person}{Melanie Mitchell}.}
  \bibinfo{year}{1998}\natexlab{}.
\newblock \bibinfo{booktitle}{\emph{An Introduction to Genetic Algorithms}}.
\newblock \bibinfo{publisher}{MIT Press}, \bibinfo{address}{Cambridge, MA,
  USA}.
\newblock
\showISBNx{0262631857}


\bibitem[\protect\citeauthoryear{{Monmarche}, {Nocent}, {Slimane}, {Venturini},
  and {Santini}}{{Monmarche} et~al\mbox{.}}{1999}]%
        {monmarche1999imagine}
\bibfield{author}{\bibinfo{person}{N. {Monmarche}}, \bibinfo{person}{G.
  {Nocent}}, \bibinfo{person}{M. {Slimane}}, \bibinfo{person}{G. {Venturini}},
  {and} \bibinfo{person}{P. {Santini}}.} \bibinfo{year}{1999}\natexlab{}.
\newblock \showarticletitle{Imagine: a tool for generating HTML style sheets
  with an interactive genetic algorithm based on genes frequencies}. In
  \bibinfo{booktitle}{\emph{IEEE SMC'99 Conference Proceedings. 1999 IEEE
  International Conference on Systems, Man, and Cybernetics (Cat.
  No.99CH37028)}}, Vol.~\bibinfo{volume}{3}. \bibinfo{publisher}{IEEE},
  \bibinfo{address}{New York, NY, USA}, \bibinfo{pages}{640--645 vol.3}.
\newblock
\urldef\tempurl%
\url{https://doi.org/10.1109/ICSMC.1999.823287}
\showDOI{\tempurl}


\bibitem[\protect\citeauthoryear{Oppenlaender, Tiropanis, and
  Hosio}{Oppenlaender et~al\mbox{.}}{2020}]%
        {oppenlaender2020crowdui}
\bibfield{author}{\bibinfo{person}{Jonas Oppenlaender},
  \bibinfo{person}{Thanassis Tiropanis}, {and} \bibinfo{person}{Simo Hosio}.}
  \bibinfo{year}{2020}\natexlab{}.
\newblock \showarticletitle{CrowdUI: Supporting Web Design with the Crowd}.
\newblock \bibinfo{journal}{\emph{Proceedings of the ACM on Human-Computer
  Interaction}} \bibinfo{volume}{4}, \bibinfo{number}{EICS}
  (\bibinfo{year}{2020}), \bibinfo{pages}{1--28}.
\newblock


\bibitem[\protect\citeauthoryear{Park, Son, Lee, and Bae}{Park
  et~al\mbox{.}}{2013}]%
        {park2013crowd}
\bibfield{author}{\bibinfo{person}{Cheong~Ha Park}, \bibinfo{person}{KyoungHee
  Son}, \bibinfo{person}{Joon~Hyub Lee}, {and} \bibinfo{person}{Seok-Hyung
  Bae}.} \bibinfo{year}{2013}\natexlab{}.
\newblock \showarticletitle{Crowd vs. Crowd: Large-Scale Cooperative Design
  through Open Team Competition}. In \bibinfo{booktitle}{\emph{Proceedings of
  the 2013 Conference on Computer Supported Cooperative Work}} (San Antonio,
  Texas, USA) \emph{(\bibinfo{series}{CSCW '13})}.
  \bibinfo{publisher}{Association for Computing Machinery},
  \bibinfo{address}{New York, NY, USA}, \bibinfo{pages}{1275–1284}.
\newblock
\showISBNx{9781450313315}
\urldef\tempurl%
\url{https://doi.org/10.1145/2441776.2441920}
\showDOI{\tempurl}


\bibitem[\protect\citeauthoryear{Quiroz, Louis, Shankar, and Dascalu}{Quiroz
  et~al\mbox{.}}{2007}]%
        {quiroz2007interactive}
\bibfield{author}{\bibinfo{person}{Juan~C Quiroz}, \bibinfo{person}{Sushil~J
  Louis}, \bibinfo{person}{Anil Shankar}, {and} \bibinfo{person}{Sergiu~M
  Dascalu}.} \bibinfo{year}{2007}\natexlab{}.
\newblock \showarticletitle{Interactive genetic algorithms for user interface
  design}. In \bibinfo{booktitle}{\emph{2007 IEEE congress on evolutionary
  computation}}. \bibinfo{publisher}{IEEE}, \bibinfo{address}{New York, NY,
  USA}, \bibinfo{pages}{1366--1373}.
\newblock


\bibitem[\protect\citeauthoryear{Reinecke and Gajos}{Reinecke and
  Gajos}{2014}]%
        {reinecke2014quantifying}
\bibfield{author}{\bibinfo{person}{Katharina Reinecke} {and}
  \bibinfo{person}{Krzysztof~Z. Gajos}.} \bibinfo{year}{2014}\natexlab{}.
\newblock \showarticletitle{Quantifying Visual Preferences around the World}.
  In \bibinfo{booktitle}{\emph{Proceedings of the SIGCHI Conference on Human
  Factors in Computing Systems}} (Toronto, Ontario, Canada)
  \emph{(\bibinfo{series}{CHI '14})}. \bibinfo{publisher}{Association for
  Computing Machinery}, \bibinfo{address}{New York, NY, USA},
  \bibinfo{pages}{11–20}.
\newblock
\showISBNx{9781450324731}
\urldef\tempurl%
\url{https://doi.org/10.1145/2556288.2557052}
\showDOI{\tempurl}


\bibitem[\protect\citeauthoryear{Rochelle~King}{Rochelle~King}{2017}]%
        {ab_oreilly}
\bibfield{author}{\bibinfo{person}{Caitlin~Tan Rochelle~King, Elizabeth
  F~Churchill}.} \bibinfo{year}{2017}\natexlab{}.
\newblock \bibinfo{booktitle}{\emph{Designing with Data}}.
\newblock \bibinfo{publisher}{O'Reilly Media, Inc.},
  \bibinfo{address}{Sebastopol, CA, USA}.
\newblock


\bibitem[\protect\citeauthoryear{Salem}{Salem}{2017}]%
        {salem2017user}
\bibfield{author}{\bibinfo{person}{Paulo Salem}.}
  \bibinfo{year}{2017}\natexlab{}.
\newblock \showarticletitle{User interface optimization using genetic
  programming with an application to landing pages}.
\newblock \bibinfo{journal}{\emph{Proceedings of the ACM on Human-Computer
  Interaction}} \bibinfo{volume}{1}, \bibinfo{number}{EICS}
  (\bibinfo{year}{2017}), \bibinfo{pages}{1--17}.
\newblock


\bibitem[\protect\citeauthoryear{Storn}{Storn}{1996}]%
        {storn1996usage}
\bibfield{author}{\bibinfo{person}{Rainer Storn}.}
  \bibinfo{year}{1996}\natexlab{}.
\newblock \showarticletitle{On the usage of differential evolution for function
  optimization}. In \bibinfo{booktitle}{\emph{Proceedings of North American
  Fuzzy Information Processing}}. \bibinfo{publisher}{IEEE},
  \bibinfo{address}{New York, NY, USA}, \bibinfo{pages}{519--523}.
\newblock


\bibitem[\protect\citeauthoryear{Takagi}{Takagi}{2001}]%
        {takagi2001interactive}
\bibfield{author}{\bibinfo{person}{Hideyuki Takagi}.}
  \bibinfo{year}{2001}\natexlab{}.
\newblock \showarticletitle{Interactive evolutionary computation: Fusion of the
  capabilities of EC optimization and human evaluation}.
\newblock \bibinfo{journal}{\emph{Proc. IEEE}} \bibinfo{volume}{89},
  \bibinfo{number}{9} (\bibinfo{year}{2001}), \bibinfo{pages}{1275--1296}.
\newblock


\bibitem[\protect\citeauthoryear{Tamburrelli and Margara}{Tamburrelli and
  Margara}{2014}]%
        {tamburrelli2014towards}
\bibfield{author}{\bibinfo{person}{Giordano Tamburrelli} {and}
  \bibinfo{person}{Alessandro Margara}.} \bibinfo{year}{2014}\natexlab{}.
\newblock \showarticletitle{Towards Automated A/B Testing}. In
  \bibinfo{booktitle}{\emph{Search-Based Software Engineering}}.
  \bibinfo{publisher}{Springer International Publishing},
  \bibinfo{address}{Cham, Switzerland}, \bibinfo{pages}{184--198}.
\newblock
\showISBNx{978-3-319-09940-8}


\bibitem[\protect\citeauthoryear{Vanderdonckt, Zen, and Vatavu}{Vanderdonckt
  et~al\mbox{.}}{2019}]%
        {vanderdonckt2019ab4web}
\bibfield{author}{\bibinfo{person}{Jean Vanderdonckt}, \bibinfo{person}{Mathieu
  Zen}, {and} \bibinfo{person}{Radu-Daniel Vatavu}.}
  \bibinfo{year}{2019}\natexlab{}.
\newblock \showarticletitle{AB4Web: An on-line A/B tester for comparing user
  interface design alternatives}.
\newblock \bibinfo{journal}{\emph{Proceedings of the ACM on Human-Computer
  Interaction}} \bibinfo{volume}{3}, \bibinfo{number}{EICS}
  (\bibinfo{year}{2019}), \bibinfo{pages}{1--28}.
\newblock


\bibitem[\protect\citeauthoryear{Vliegendhart, Dolstra, and
  Pouwelse}{Vliegendhart et~al\mbox{.}}{2012}]%
        {vliegendhart2012crowdsourced}
\bibfield{author}{\bibinfo{person}{Raynor Vliegendhart}, \bibinfo{person}{Eelco
  Dolstra}, {and} \bibinfo{person}{Johan Pouwelse}.}
  \bibinfo{year}{2012}\natexlab{}.
\newblock \showarticletitle{Crowdsourced User Interface Testing for Multimedia
  Applications}. In \bibinfo{booktitle}{\emph{Proceedings of the ACM Multimedia
  2012 Workshop on Crowdsourcing for Multimedia}} (Nara, Japan)
  \emph{(\bibinfo{series}{CrowdMM '12})}. \bibinfo{publisher}{Association for
  Computing Machinery}, \bibinfo{address}{New York, NY, USA},
  \bibinfo{pages}{21–22}.
\newblock
\showISBNx{9781450315890}
\urldef\tempurl%
\url{https://doi.org/10.1145/2390803.2390813}
\showDOI{\tempurl}


\bibitem[\protect\citeauthoryear{Wang, Varvello, and Kuzmanovic}{Wang
  et~al\mbox{.}}{2019}]%
        {wang2019kaleidoscope}
\bibfield{author}{\bibinfo{person}{Pengfei Wang}, \bibinfo{person}{Matteo
  Varvello}, {and} \bibinfo{person}{Aleksandar Kuzmanovic}.}
  \bibinfo{year}{2019}\natexlab{}.
\newblock \showarticletitle{Kaleidoscope: A crowdsourcing testing tool for web
  quality of experience}. In \bibinfo{booktitle}{\emph{2019 IEEE 39th
  International Conference on Distributed Computing Systems (ICDCS)}}.
  \bibinfo{publisher}{IEEE}, \bibinfo{address}{New York, NY, USA},
  \bibinfo{pages}{1971--1982}.
\newblock


\bibitem[\protect\citeauthoryear{Xu, Huang, and Bailey}{Xu
  et~al\mbox{.}}{2014}]%
        {xu2014voyant}
\bibfield{author}{\bibinfo{person}{Anbang Xu}, \bibinfo{person}{Shih-Wen
  Huang}, {and} \bibinfo{person}{Brian Bailey}.}
  \bibinfo{year}{2014}\natexlab{}.
\newblock \showarticletitle{Voyant: Generating Structured Feedback on Visual
  Designs Using a Crowd of Non-Experts}. In
  \bibinfo{booktitle}{\emph{Proceedings of the 17th ACM Conference on Computer
  Supported Cooperative Work \& Social Computing}} (Baltimore, Maryland, USA)
  \emph{(\bibinfo{series}{CSCW '14})}. \bibinfo{publisher}{Association for
  Computing Machinery}, \bibinfo{address}{New York, NY, USA},
  \bibinfo{pages}{1433–1444}.
\newblock
\showISBNx{9781450325400}
\urldef\tempurl%
\url{https://doi.org/10.1145/2531602.2531604}
\showDOI{\tempurl}


\end{thebibliography}

\end{document}